\documentclass[usenatbib,referee]{mn2e}
\usepackage{amsmath}
\usepackage{graphicx}
\usepackage{epsfig}
\usepackage{txfonts}
\usepackage{color}
\usepackage{natbib}

\title[Off-centre carbon burning in He-accreting CO WDs]
{Off-centre carbon burning in He-accreting carbon-oxygen white dwarfs}
\author[C. Wu, B. Wang]
{Chengyuan Wu$^{\rm 1,2,3}$\thanks{E-mail:wcy@ynao.ac.cn}, Bo Wang$^{\rm 1,2,3}$\thanks{E-mail:wangbo@ynao.ac.cn}\\
$^1$Key Laboratory for the Structure and Evolution of Celestial Objects, Yunnan Observatories, CAS, Kunming 650216, China\\
$^2$University of Chinese Academy of Sciences, Beijing 100049, China\\
$^3$Centre for Astronomical Mega-Science, CAS, Beijing 100012, China}

\begin{document}
%\date{Accepted. Received}
\date{}
\pagerange{\pageref{firstpage}--\pageref{lastpage}} \pubyear{2018}
\maketitle

\label{firstpage}

\begin{abstract}\label{0. abstract}
The carbon-oxygen white dwarf (CO WD) + He star channel is one of the promising ways for producing type Ia supernovae (SNe Ia) with short delay times. Recent studies found that carbon under the He-shell can be ignited if the mass-accretion rate of CO WD is higher than a critical rate (about $2\times10^{-6}\,{M}_\odot\,\mbox{yr}^{-1}$), triggering an inwardly propagating carbon flame. Previous studies usually supposed that the off-centre carbon flame would reach the centre, resulting in the formation of an oxygen-neon (ONe) WD that will collapse into a neutron star. However, the process of off-centre carbon burning is not well studied. This may result in some uncertainties on the final fates of CO WDs. By employing MESA, we simulated the long-term evolution of off-centre carbon burning in He-accreting CO WDs. We found that the inwardly propagating carbon flame transforms the CO WDs into OSi cores directly but not ONe cores owing to the high temperature of the burning front. We suggest that the final fates of the CO WDs may be OSi WDs under the conditions of off-centre carbon burning, or explode as iron-core-collapse SNe if the mass-accretion continues. We also found that the mass-fractions of silicon in the OSi cores are sensitive to the mass-accretion rates.
\end{abstract}


\section{Introduction} \label{1. Introduction}

Type Ia supernovae (SNe Ia) are used as distance indicators successfully to determine the cosmological parameters (e.g. Riess et al. 1998; Perlmutter et al. 1999) and play a key role in the chemical evolution of galaxies (e.g. Greggio \& Renzini 1983; Matteucci \& Greggio 1986). They are suggested to originate from thermonuclear runaway of carbon-oxygen white dwarfs (CO WDs) when the WDs grow in mass approaching the Chandrasekhar-mass limit (${M}_{\rm Ch}$; e.g. Hoyle \& Fowler 1960; Woosley, Taam \& Weaver 1986).

So far, the single-degenerate model and the double-degenerate model are two popular progenitor models for the formation of SNe Ia, which have been discussed for many decades (e.g. Whelan \& Iben 1973; Webbink 1984). In the single-degenerate model, a CO WD accretes H- or He-rich material from its non-degenerate companion (e.g. a main-sequence star, a red-giant star or a He star) and finally an SN Ia may be produced if the WD grows in mass close to ${M}_{\rm Ch}$ (e.g. Hachisu, Kato \& Nomoto 1996; Li \& van den Heuvel 1997; Han \& Podsiadlowski 2004; Wang 2018). In the double-degenerate model, an SN Ia may be produced if the total mass of the double WDs exceeds ${M}_{\rm Ch}$ (e.g. Iben \& Tutukov 1984).

In the single-degenerate model, CO WD+He star systems are thought to be one of the promising ways for producing SNe Ia with short delay times (e.g. Wang et al. 2009; Ruiter, Belczynski \& Fryer 2009; Liu et al. 2010). The He-rich material is transferred from the companion star on to the surface of the CO WD, and burns into carbon and oxygen, resulting in the mass-increase of the WD. During the mass-transfer process, the mass-accretion rate ($\dot{M}_{\rm acc}$) is a key parameter for the evolution of CO WDs. The accumulated He-rich material can burn into carbon and oxygen stably if $\dot{M}_{\rm acc}$ is in the steady burning region. In this case, the CO WD can grow in mass steadily to ${M}_{\rm Ch}$ and explodes as an SN Ia (e.g. Piersanti, Tornamb\'{e}, \& Yungelson 2014; Wang et al. 2015; Wu et al. 2016).

Wang, Podsiadlowski \& Han (2017) found that there exists a critical mass-accretion rate ($\dot{M}_{\rm cr}$) in the steady burning region, and carbon under the He-shell ignites before the mass of the WD reaches ${M}_{\rm Ch}$ if $\dot{M}_{\rm acc}>\dot{M}_{\rm cr}$ (see also Saio \& Nomoto 1998; Brooks et al. 2016). Previous studies usually assumed that the inwardly propagating carbon flames transform the CO WDs into ONe cores, and subsequently the ONe cores may collapse into neutron stars (NSs) by the electron-capture of $^{\rm 24}{\rm Mg}$ and $^{\rm 20}{\rm Ne}$ if the ONe cores can increase their masses continuously (e.g. Saio \& Nomoto 1985, 1998; Nomoto \& Kondo 1991; Brooks et al. 2017; Liu \& Li 2017). However, in the previous studies the inwardly propagating carbon flames did not reach the centre of the WDs, leading to some uncertainties on the final fates of the He-accreting CO WDs.

In this article, we aim to simulate the long-term evolution of off-centre carbon burning in He-accreting CO WDs and to investigate the final outcomes of the CO WDs. In Sect. 2, we provide our basic assumptions and methods for the numerical simulations. The results of our simulations are shown in Sect. 3. Finally, we present discussion and conclusions in Sect. 4.

\section{Numerical Methods}\label{Methods}

We use the stellar evolution code \texttt{MESA} (version 7624; see Paxton et al. 2011, 2013, 2015) to simulate the off-centre carbon burning in He-accreting CO WDs. The OPAL opacity is adopted (e.g. Iglesias \& Rogers 1996). This is applicable to extra carbon and oxygen during the He burning. We adopted \texttt{co\_burn.net} as the nuclear reaction network in our simulations. This nuclear reaction network includes the isotopes needed for helium, carbon and oxygen burning (i.e., $^{\rm 3}{\rm He}$, $^{\rm 4}{\rm He}$, $^{\rm 7}{\rm Li}$, $^{\rm 7}{\rm Be}$, $^{\rm 8}{\rm B}$, $^{\rm 12}{\rm C}$, $^{\rm 14}{\rm N}$, $^{\rm 15}{\rm N}$, $^{\rm 16}{\rm O}$, $^{\rm 19}{\rm F}$, $^{\rm 20}{\rm Ne}$, $^{\rm 23}{\rm Na}$, $^{\rm 24}{\rm Mg}$, $^{\rm 27}{\rm Al}$ and $^{\rm 28}{\rm Si}$), which are coupled by 57 nuclear reactions.

First, we construct a hot CO core with initial mass of ${M}_{\rm WD}^{\rm i}=0.9{M}_\odot$. In our simulations, the hot CO core is built in idealized methods, in which the mass-fractions of $^{\rm 12}{\rm C}$ and $^{\rm 16}{\rm O}$ are both $50\%$ (e.g. Saio \& Nomoto 1998; Schwab, Quataert \& Kasen 2016). During the formation process, all nuclear reactions are neglected, resulting in an uniform elemental abundance distribution of the core from the inside out. Then, we cool down the CO core for a given time of ${t}_{\rm cool}=10^{6}\,\mbox{yr}$. At this moment, the initial WD model has been constructed, in which the effective temperature is ${\rm {log}_{10}}({T}_{\rm eff}/{\rm K})=5.367$ and the central temperature is ${\rm {log}_{10}}({T}_{\rm c}/{\rm K})=7.865$.

Secondly, we simulate the long-term evolution of He-accreting CO WDs. The mass-fraction of He and metallicity of the accreted material is set to be 0.98 and 0.02, respectively. We use a constant mass-accretion rate of $\dot{M}_{\rm acc}=4\times10^{-6}\,{M}_\odot\,\mbox{yr}^{-1}$ in our simulations, in which the WD undergoes stable He-shell burning. At beginning of the He-shell burning, the energy release rate is relatively high, resulting in a He-shell flash. We adopted the super-Eddington wind as the mass-loss mechanism during the first He-shell flash (see also Denissenkov et al. 2013a; Wu et al. 2017; Wu \& Wang 2018). In the super-Eddington wind assumption, if the luminosity ($L_{\rm eff}$) on the surface of the WD exceeds the Eddington luminosity ($L_{\rm Edd}$), the super-Eddington wind would be triggered and blow away part of the accumulated material outside the CO core.

\section{Numerical Results}\label{Results}

Figs\,1 to 3 show the results of off-centre carbon burning, in which ${M}_{\rm WD}^{\rm i}=0.9{M}_\odot$ and $\dot{M}_{\rm acc}=4.0\times10^{-6}\,{M}_\odot\,\mbox{yr}^{-1}$. The accreted He-rich material burns into carbon and oxygen, resulting in the mass-increase of the CO WD, and the surface of CO core under the He-shell is ignited when the core mass increases to about $1.277{M}_\odot$.

\begin{figure}
\begin{center}
\includegraphics[width=0.7\textwidth]{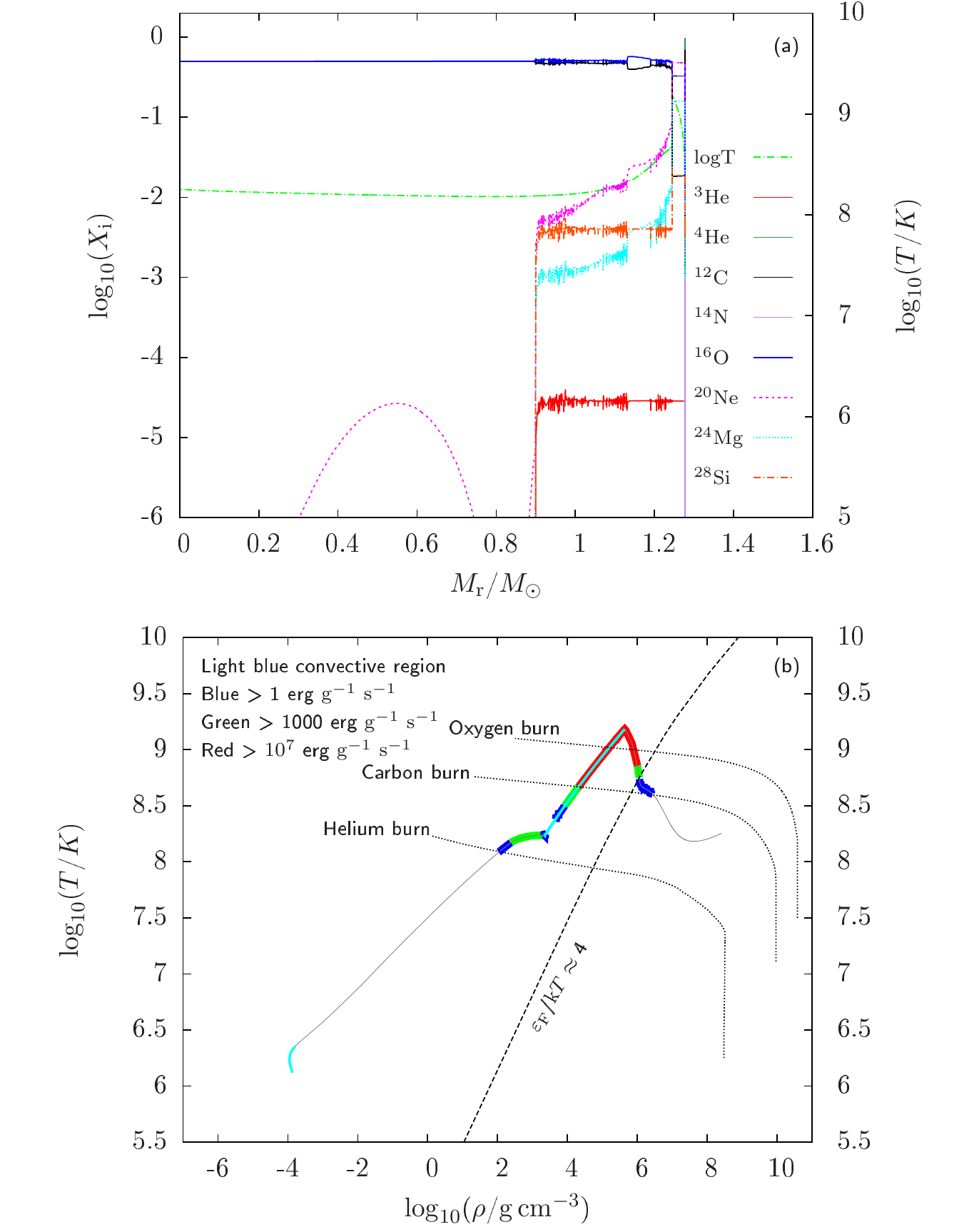}
 \caption{Results of off-centre carbon burning at the moment when carbon under the He-shell ignites, in which ${M}_{\rm WD}^{\rm i}=0.9{M}_\odot$ and $\dot{M}_{\rm acc}=4\times10^{-6}\,{M}_\odot\,\mbox{yr}^{-1}$. Panel (a): the elemental abundance distribution profile. Panel (b): the density-temperature profile, in which the dashed line shows the separation of degenerate and non-degenerate regions (${\varepsilon}_{\rm F}/{\rm k}{T}\approx4$, where ${\varepsilon}_{\rm F}$ is the Fermi energy), light blue curves indicates the convective region, the grey curve signifies the non-convective region, the blue, green and red curves show the energy production rates of nuclear reactions, respectively.}
  \end{center}
\end{figure}

Fig.\,1 presents the abundance distribution and the density-temperature profile of the CO core at the moment when the carbon under the He-shell is ignited. The carbon burning produces an outwardly extending convective zone, leading to the expansion of the star. However, the convective zone does not extend to the outside convective zone produced by He-shell burning and subsequently the He-shell burning begins to quench. As the burning releases energy outwardly, the carbon burning zone shrinks gradually. At the same time, the off-centre carbon flame begins to propagate inward.

\begin{figure}
\begin{center}
\includegraphics[width=0.7\textwidth]{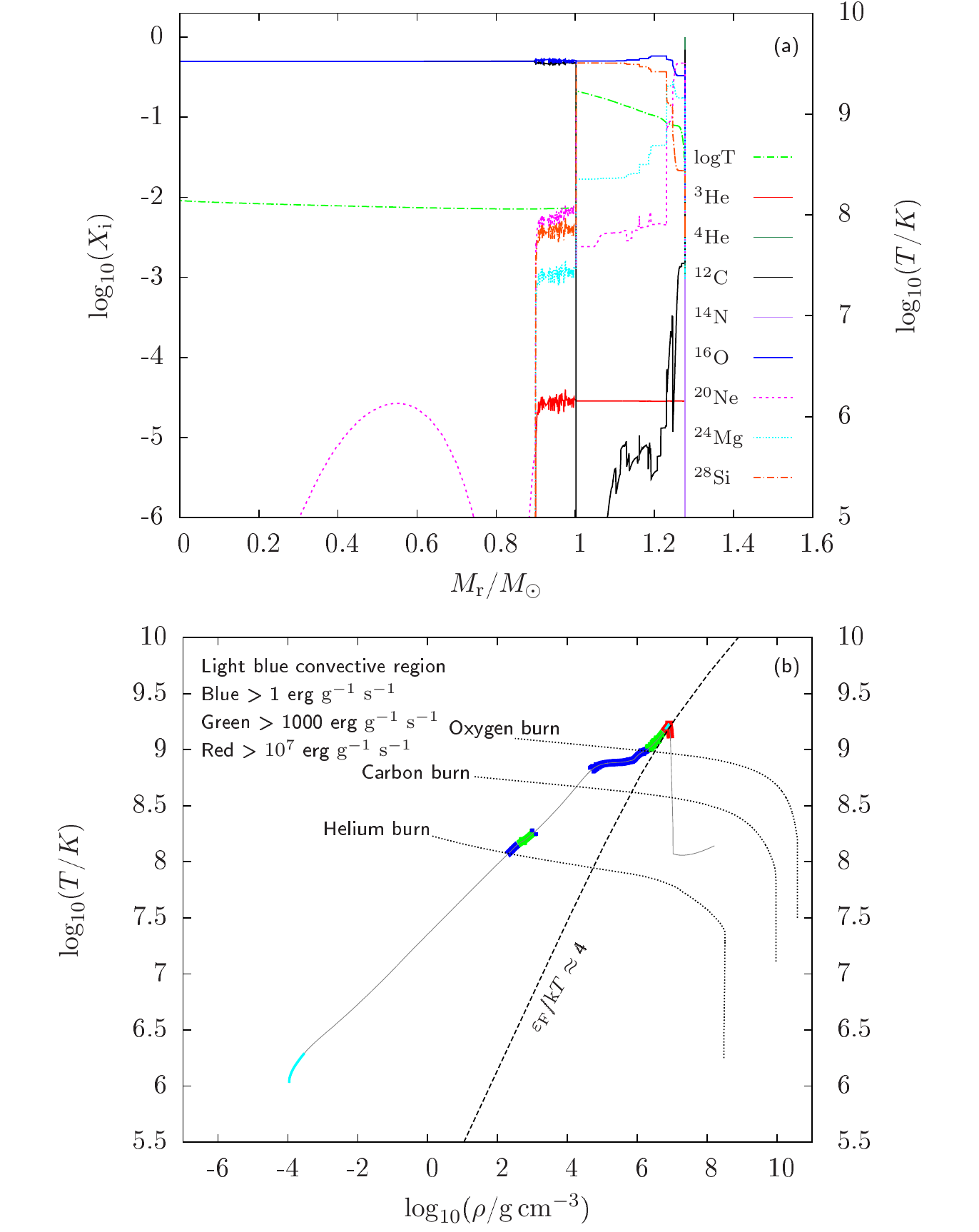}
 \caption{Similar to Fig.\,1, but for the moment when the off-centre carbon burning flame reached the position where the mass-coordinate ${M}_{\rm r}=1.0{M}_\odot$.}
  \end{center}
\end{figure}

Owing to the high temperature of the carbon flame, $^{\rm 20}{\rm Ne}$ is transformed into $^{\rm 28}{\rm Si}$ once it appears, resulting in an extremely thin neon flame (about $1\,\mbox{km}$ thick). Fig.\,2 shows the abundance distribution and the density-temperature profile of the CO core when the carbon burning flame reaches the position where the mass-coordinate ${M}_{\rm r}\approx1{M}_\odot$. At this moment, the abundance of $^{\rm 16}{\rm O}$ and $^{\rm 28}{\rm Si}$ near the flame are almost the same.

\begin{figure}
\begin{center}
\includegraphics[width=0.7\textwidth]{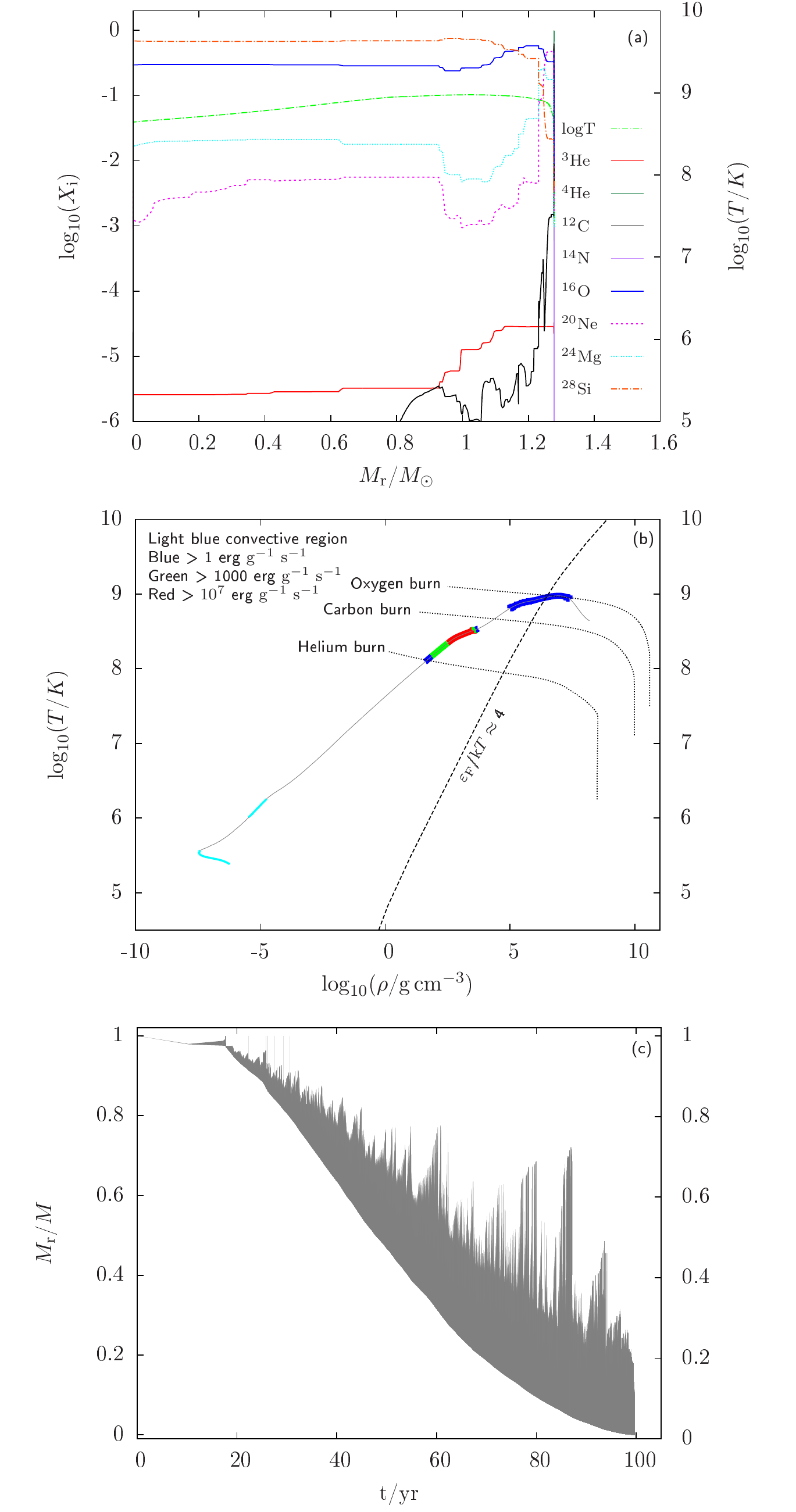}
 \caption{Similar to Fig.\,1, but for the moment after the off-centre carbon burning flame reached the centre. Panel (c): The kippenhahn diagram during the carbon flame propagating phase, in which x-axis represents evolution time and y-axis is mass-coordinate.}
  \end{center}
\end{figure}

As the carbon burning flame keeps propagating inward, the density of the burning region becomes higher, and the nuclear reaction rate increases at the same time, resulting in more $^{\rm 28}{\rm Si}$ produced. In Fig.\,3, we present the abundance distribution (panel a) and the density-temperature profile (panel b) of the CO core when the carbon flame reached the centre. From this figure, we can see that the core is transformed into an OSi core which is surrounded by an ONe-shell and a thin He-rich envelope. It takes about one hundred years for the carbon flame to propagate from the surface to the centre. The dynamical time-scale of the WD is $\tau_{\rm dyn}\approx({R}^{3}/{\rm G}M)^{0.5}\approx2.7\,\mbox{s}$ in this example, whereas the flame propagating time-scale ${\tau}_{\rm f}\approx100\,\mbox{yr}$. This indicates that ${\tau}_{\rm f}\gg{\tau}_{\rm dyn}$ and the flame propagating process does not require special hydrodynamic treatment. The Kippenhahn diagram during the flame propagating phase is shown in panel (c). From this figure we can see that the carbon flame moves inwardly in an unstable way, resulting in the fluctuations of the elemental abundance distributions of carbon, neon and magnesium in panel (a). After the burning front reaches the centre, the carbon flame quenches gradually and the OSi core begins to cool down at the same time. On the surface of the OSi core, the He-shell is reignited owing to the continuation of mass-accretion. As a result, the OSi core reaccumulates a CO-shell until next carbon ignition occurs. The final outcomes of the CO core can be divided into two cases as follows: (1) If the OSi core is not massive enough to trigger an off-centre silicon burning, the core would cool down to form a massive OSi WD. This implies a new class of WDs may exist in the Universe. (2) If the WD accretes material continuously, the off-centre silicon ignition may be triggered and transforms silicon into iron. Finally, a NS would be formed through the iron-core-collapse SN (Fe-CCSN), likely relating to the formation of some faint SNe. The present work may increase the birthrates of OSi WDs and Fe-CCSNe in the Universe.

\section{Discussion and conclusions}\label{Discussion and conclusions}

\begin{figure}
\begin{center}
\includegraphics[width=0.7\textwidth]{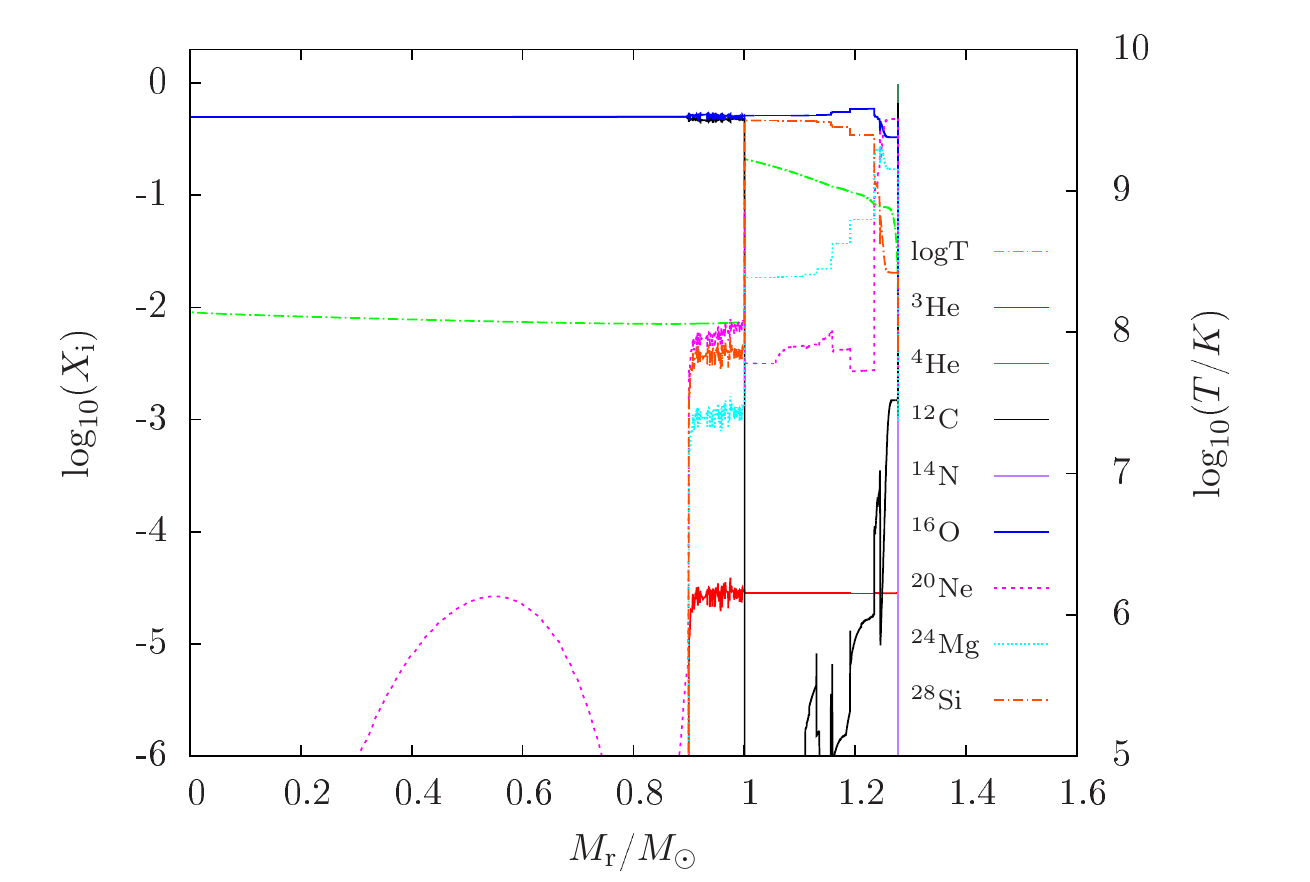}
 \caption{Similar to panel (a) of Fig.\,2, but for the increase of temporal and spatial resolution.}
  \end{center}
\end{figure}

It is difficult to resolve the flame that is propagating inside the star, and the computational results may be sensitive to the temporal and spatial resolution. We limited the time-step of each model and increased the mass of zones by decreasing the value of ``varcontrol targer'' and ``mesh delta coeff''. We made several convergence tests to confirm the fact that our results are independent of the choice of time-step and zoning. In Fig.\,4, we replot the elemental abundance distribution profile when the carbon burning flame propagates to the position where the mass-coordinate ${M}_{\rm r}=1.0{M}_\odot$ (similar to panel (a) of Fig.\,2) by both limiting the time-step and increasing the number of mass zone. In this case, the number of mass zone is increased from about $1700$ to more than $3500$. From this figure, we can see that the elemental abundances are similar to those in Fig.\,2. Our tests indicate that the present results are indeed numerically converged.

\begin{figure}
\begin{center}
\includegraphics[width=0.7\textwidth]{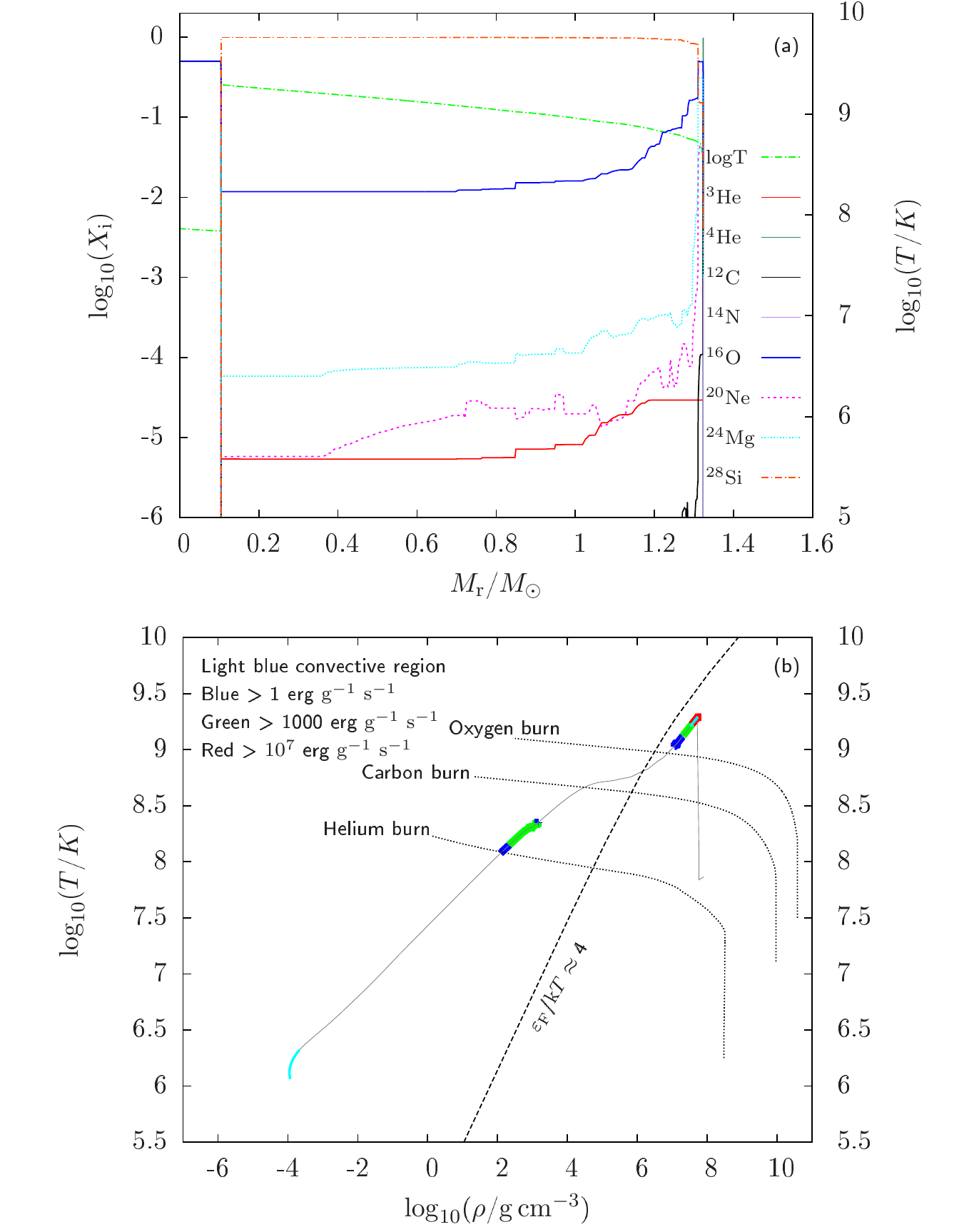}
 \caption{Results of off-centre carbon burning, in which ${M}_{\rm WD}^{\rm i}=0.9{M}_\odot$ and $\dot{M}_{\rm acc}=3\times10^{-6}\,{M}_\odot\,\mbox{yr}^{-1}$. In this case, the inwardly propagating carbon flame reaches the position where ${M}_{\rm r}=0.1{M}_\odot$. Panel (a): the elemental abundance distribution profile. Panel (b): the density-temperature profile.}
  \end{center}
\end{figure}

In our simulations, we set the mass-accretion rate as $4\times10^{-6}\,{M}_\odot\,\mbox{yr}^{-1}$. However, $\dot{M}_{\rm acc}$ may have influence on the final outcomes of the He-accreting WDs. Here, we use ${M}_{\rm WD}^{\rm i}=0.9{M}_\odot$ and $\dot{M}_{\rm acc}=3\times10^{-6}\,{M}_\odot\,\mbox{yr}^{-1}$ as a comparison. In Fig.\,5, we present the abundance distribution and the density-temperature profile of the $0.9{M}_\odot$ CO WD when the inwardly propagating carbon flame reached the position where the mass-coordinate ${M}_{\rm r}=0.1{M}_\odot$. The WD increases its mass to $1.323{M}_\odot$ when the off-centre carbon is ignited, in which the mass of the core is higher than the case in Sect.\,3. From this figure, we can see that the mass-fraction of $^{\rm 28}{\rm Si}$ produced by carbon and neon burning becomes higher as $\dot{M}_{\rm acc}$ decreases. This is because the degeneracy of the CO-shell grows as $\dot{M}_{\rm acc}$ decreases, resulting in the higher temperature of burning front. In this case, we stop our calculation when the burning wave reaches the position where ${M}_{\rm r}=0.05{M}_\odot$, because the calculation becomes extremely time-consuming when the flame propagates to the innermost part of the core. The inwardly propagating carbon flame would finally reach the centre, transforming the CO WD into a Si core. The final fate of the core would be similar to the case in Sect.\,3. For the lower mass-accretion rate (e.g. we simulated the evolution of the accreting WD by using $\dot{M}_{\rm acc}=2\times10^{-6}\,{M}_\odot\,\mbox{yr}^{-1}$), a rapidly outwardly extending convective zone caused by carbon burning would extend to the He-burning shell once the off-centre carbon is ignited, resulting in a large energy release. Subsequently, the gravitational binding energy may not bound such a high energy generation rate (the surface velocity in such a simulation becomes extremely high), likely triggering a dynamical process along with a strong mass-loss on the surface of the WD. To understand the final outcomes of the He-accreting WDs under this condition, further dynamical simulations are needed.

Besides the mass-accretion rate, the initial mass of the WD may also be important for the final fates of He-accreting CO WDs. Kato, Saio \& Hachisu (2018) investigated the evolution of He-accreting CO WDs that experience He-flashes. They found that the mass-fraction of $^{\rm 28}{\rm Si}$ becomes obviously higher in He-burning ashes for more massive initial WDs. For the present work, the mass of initial WD may also influence the evolution of off-centre carbon burning. For a massive WD, the temperature of accumulated He-shell is higher owing to the high surface gravitational acceleration, leading to a thinner He-shell when carbon is ignited. The inwardly propagating carbon flame takes more time to reach the centre and the mass-fraction of $^{\rm 28}{\rm Si}$ may become higher in carbon burning ashes owing to the high density of the core.

In our simulations, we did not consider the convective boundary mixing. Denissenkov et al. (2013a,b) investigated the inwardly propagating carbon flame in the CO cores by considering a convective boundary mixing assumption, and found that the mixing effect can prevent the carbon flames from reaching the centre, resulting in the formation of hybrid CONe WDs (see also Farmer, Fields \& Timmes 2015). If the convective boundary mixing is considered in our simulations, the final structure of the CO WD may be a COSi core surrounded by an ONe-shell and a thin He-rich envelope.

HD 49798/RX J0648.0-4418 (e.g. Mereghetti et al. 2011) is a binary system consisting of a H depleted subdwarf star and a compact companion. Recent studies suggested that the compact star in this system is a WD rather than a NS (e.g. Mereghetti et al. 2013; Liu et al. 2015, 2018). The WD (${M}_{\rm WD}=1.28\pm0.05{M}_\odot$) is accreting He-rich material from its companion (${M}_{\rm He}=1.50\pm0.05{M}_\odot$) through the stellar wind at a rate of $10^{-11}<\dot{M}_{\rm acc}<10^{-10}\,{M}_\odot\,\mbox{yr}^{-1}$ (e.g. Bisscheroux et al. 1997; Mereghetti et al. 2009). Wang \& Han (2010) estimated that the He star will fill its Roche lobe in about $4\times10^{4}\,\mbox{yr}$, and the WD will grow to ${M}_{\rm Ch}$. Brooks et al. (2017) simulated the binary evolution of ONe WD+He star system, in which $1.1\leq{M}_{\rm WD}\leq1.3{M}_\odot$, ${M}_{\rm He}=1.5{M}_\odot$ and $3\,\mbox{hr}\leq{P}_{\rm orb}\leq8\,\mbox{d}$, and found that the mass-accretion rate of the WD is higher than $3\times10^{-6}\,{M}_\odot\,\mbox{yr}^{-1}$ when the He donor fills its Roche lobe. According to our present work, RX J0648.0-4418 may experience off-centre carbon burning and then form a NS through Fe-CCSN if the compact star is a CO WD. Thus, we suggest that the final fate of HD 49798/RX J0648.0-4418 will be a NS system whether the primary star is a WD (through Fe-CCSN if it is a CO WD or through electron capture SN if it is a ONe WD) or a NS (see also Brooks, Kupfer \& Bildsten 2017).

The OSi WDs may also be produced in the double-degenerate model. Some recent studies suggested that the off-centre carbon burning would occur during the merging of two CO WDs, and the inwardly propagating flame may transform the CO WD into a Si core if the merge of double CO WDs satisfies some conditions. For example, the mass of the merger remnants is massive enough (see Schwab, Quataert \& Kasen 2016) or the mass-accretion rate during the merger process is in a certain limits (see Wu, Wang \& Liu 2019). Eventually, the Si core may cool down to form a Si WD if the off-centre Si burning does not occur. Alternatively, the Si core may form a NS through Fe-CCSN if the core mass is high enough to form an iron core.

By employing the stellar evolution code MESA, we investigated the long-term evolution of off-centre carbon burning of CO WDs by accreting He-rich material. We found that the off-centre carbon burning transforms CO WDs into OSi cores, in which the mass-fraction of Si is sensitive to the mass accretion rate, i.e. the lower $\dot{M}_{\rm acc}$ results in the higher mass-fraction of Si. The present work indicates that the CO WDs accreting He-rich material will form OSi WDs or experience Fe-CCSNe rather than ECSNe if the off-centre carbon ignition occurs. The present work increases the birthrates of OSi WDs and Fe-CCSNe in the Universe. In order to further understand the evolutions of He-accreting WDs, more CO WD+He star systems are hoped to be detected, and more numerical simulations on off-centre carbon burning are needed.

\section*{Acknowledgments}

We acknowledge the referee, Christopher Tout, for the valuable comments that helped us to improve the paper.
This study is supported by the National Natural Science Foundation of China (Nos 11873085, 11673059 and 11521303),
the Chinese Academy of Sciences (Nos KJZD-EW-M06-01 and QYZDB-SSW-SYS001),
and the Natural Science Foundation of Yunnan Province (Nos 2018FB005 and 2017HC018).

\label{lastpage}

\begin{thebibliography}{}\label{thebibliography}%%

\bibitem[Bisscheroux et al. (1997)]{Biss97}                   Bisscheroux B. C., Pols O. R., Kahabka P., Belloni T., van den Heuvel E. P. J., 1997, A\&A, 317, 815
\bibitem[Brooks et al. (2016)]{Bro16}                         Brooks J., Bildsten L., Schwab J., Paxton B., 2016, ApJ, 821, 28
\bibitem[Brooks et al. (2017)]{Bro17}                         Brooks J., Schwab J., Bildsten L., Quataert E., Paxton B., 2017, ApJ, 843, 151
\bibitem[Brooks, Kupfer \& Bildsten (2017)]{BKB17}            Brooks J., Kupfer T., Bildsten L., 2017, ApJ, 847, 78
\bibitem[Denissenkov et al. (2013a)]{Den13a}                  Denissenkov P. A., Herwig F., Bildsten L., Paxton B., 2013a, ApJ, 762, 8
\bibitem[Denissenkov et al. (2013b)]{Den13b}                  Denissenkov P. A., Herwig F., Truran J. W., Paxton B., 2013b, ApJ, 772, 37
\bibitem[Farmer, Fields \& Timmes (2015)]{FFT15}              Farmer R., Fields C. E., Timmes F. X., 2015, ApJ, 807, 184
\bibitem[Greggio \& Renzini (1983)]{GR93}                     Greggio L., \& Renzini A., 1983, A\&A, 118, 217
\bibitem[Hachisu, Kato \& Nomoto (1996)]{HKN96}               Hachisu I., Kato M., Nomoto K., 1996, ApJ, L97
\bibitem[Han \& Podsiadlowski (2004)]{HP04}                   Han Z., \& Podsiadlowski P., 2004, MNRAS, 350, 1301
\bibitem[Hoyle \& Fowler (1960)]{HF60}                        Hoyle F., Fowler W. A., 1960, ApJ, 132, 565
\bibitem[Iben \& Tutukov (1984)]{IT84}                        Iben I., Tutukov A. V., 1984, ApJS, 54, 335
\bibitem[Iglesias \& Rogers (1996)]{IR96}                     Iglesias C. A., Rogers F. J., 1996, ApJ, 464, 943
\bibitem[Kato, Saio \& Hachisu (2018)]{KSH18}                 Kato M., Saio H., Hachisu I., 2018 ApJ, 863, 125
\bibitem[Li \& van den Heuvel (1997)]{LV97}                   Li X.-D., van den Heuvel E. P. J., 1997, A\&A, 322, L9
\bibitem[Liu \& Li (2017)]{LL17}                              Liu W., \& Li, X., 2017, ApJ, 851, 58
\bibitem[Liu et al. (2010)]{Liu10}                            Liu W., Chen W., Wang B., Han Z., 2010, A\&A, 523, A3
\bibitem[Liu et al. (2015)]{Liu15}                            Liu D., Zhou W., Wu C., Wang B., 2015, Res. Astron. Astronphys., 15, 1813
\bibitem[Liu et al. (2018)]{Liu18}                            Liu D., Chen W., Zuo Z., Han Z., 2018, MNRAS, 477, 384
\bibitem[Matteucci \& Greggio (1986)]{MG86}                   Matteucci F., \& Greggio L., 1986, A\&A, 154, 279
\bibitem[Mereghetti et al. (2009)]{Mere09}                    Mereghetti S., Tiengo A., Esposito P., et al., 2009, Science, 325, 1222
\bibitem[Mereghetti et al. (2011)]{Mere11}                    Mereghetti S., La Palombara N., Tiengo A., et al., 2011, ApJ, 737, 51
\bibitem[Mereghetti et al. (2013)]{Mere13}                    Mereghetti S., La Palombara N., Tiengo A., et al., 2013, A\&A, 533, A46
\bibitem[Nomoto \& Kondo (1991)]{NK91}                        Nomoto K., Kondo Y., 1991, ApJL, 367, L19
\bibitem[Paxton et al. (2011)]{pax11}                         Paxton B., Bildsten L., Dotter A., Herwig F., Lesaffre P., Timmes F., 2011, ApJS, 192, 3
\bibitem[Paxton et al. (2013)]{pax13}                         Paxton B., Cantiello M., Arras P., et al., 2013, ApJS, 208, 4
\bibitem[Paxton et al. (2015)]{pax15}                         Paxton B., Marchant P., Schwab J., et al., 2015, ApJS, 220, 15
\bibitem[Perlmutter et al. (1999)]{Per99}                     Perlmutter S., Aldering G., Goldhaber G., et al., 1999, ApJ, 517, 565
\bibitem[Piersanti, Tornamb\'{e} \& Yungelson (2014)]{pier14} Piersanti L., Tornamb\'{e} A., Yungelson L., 2014, MNRAS, 445, 3239
\bibitem[Riess et al. (1998)]{Rie98}                          Riess A. G., Filippenko A. V., Challis P., et al., 1998, AJ, 116, 1009
\bibitem[Ruiter, Belczynski \& Fryer (2009)]{RBF09}           Ruiter A. J., Belczynski K., Fryer C. L., 2009, ApJ, 699, 2026
\bibitem[Saio \& Nomoto (1985)]{SN85}                         Saio H., Nomoto K., 1985, A\&A, 150, L21
\bibitem[Saio \& Nomoto (1998)]{SN98}                         Saio H., Nomoto K., 1998, ApJ, 500, 388
\bibitem[Schwab, Quataert \& Kasen (2016)]{Sch16}             Schwab J., Quataert E., Kasen D., 2016, MNRAS, 463, 3461
\bibitem[Wang (2018)]{Wang18}                                 Wang B., 2018, Res. Astron. Astrophys., 18, 49
\bibitem[Wang et al. (2009)]{Wang09}                          Wang B., Meng X., Chen X., Han Z., 2009, MNRAS, 395, 847
\bibitem[Wang \& Han (2010)]{WH10}                            Wang B., Han Z., 2010, Res. Astron. Astronphys., 10, 681
\bibitem[Wang et al. (2015)]{Wang15}                          Wang B., Li Y., Ma X., et al., 2015, A\&A, 584, A37
\bibitem[Wang, Podsiadlowski \& Han (2017)]{Wang17}           Wang B., Podsiadlowski P., Han Z., 2017, MNRAS, 472, 1593
\bibitem[Webbink (1984)]{Web84}                               Webbink R. F., 1984, ApJ, 277, 355
\bibitem[Whelan \& Iben (1973)]{WI73}                         Whelan J., Iben I., 1973, ApJ, 186, 1007
\bibitem[Woosley, Taam \& Weaver (1986)]{WTW86}               Woosley S. E., Taam R. E., Weaver T. A., 1986, ApJ, 301, 601
\bibitem[Wu et al. (2016)]{Wu16}                              Wu C., Liu D., Zhou W., Wang B., 2016, Res. Astron. Astrophys., 16, 160
\bibitem[Wu et al. (2017)]{Wu17}                              Wu C., Wang B., Liu D., Han Z., 2017, A\&A, 604, A31
\bibitem[Wu \& Wang (2018)]{Wu18}                             Wu C., Wang B., 2018, Res. Astron. Astronphys., 18, 36
\bibitem[Wu, Wang \& Liu (2019)]{Wu19}                        Wu C., Wang B., Liu D., 2019, MNRAS, 483, 263

\end{thebibliography}
\end{document}